\PassOptionsToPackage{table}{xcolor}
\documentclass[10pt,final,journal,a4paper,twoside,twocolumn,romanappendices]{IEEEtran}
\IEEEoverridecommandlockouts
\usepackage[dvips]{psfrag}
\usepackage[dvips]{epsfig}
\usepackage{epic,color}
\usepackage[cmex10]{amsmath}
\usepackage{amsfonts}
\usepackage{amssymb}
\usepackage{amsmath}
\usepackage{bm}
\usepackage{yfonts}
\usepackage{array}
\usepackage{bigstrut}
\usepackage{multirow}
\usepackage{dsfont}
\usepackage{cite}
\usepackage{fixltx2e}
\usepackage{accents}
\usepackage{mathtools}
\usepackage[normalem]{ulem}
\usepackage{enumerate}
\usepackage{stfloats}
\usepackage[T1]{fontenc}
\usepackage{float}
\usepackage{epstopdf}
\usepackage{mathtools}
\usepackage{algorithm}
\usepackage{algorithmicx}
\usepackage{algpseudocode}
\usepackage{algorithm,algpseudocode,float}
\usepackage{lipsum}
\usepackage{soul,xcolor}
\usepackage{xparse}
\usepackage{acronym}
\usepackage[keeplastbox]{flushend}
\usepackage{soul}
\usepackage{enumitem}
\usepackage{float}
\usepackage{color, colortbl}
\usepackage{xcolor} 
\NewDocumentCommand{\ceil}{s O{} m}{%
  \IfBooleanTF{#1} 
    {\left\lceil#3\right\rceil} 
    {#2\lceil#3#2\rceil} 
}
\NewDocumentCommand{\floor}{s O{} m}{%
  \IfBooleanTF{#1} 
    {\left\lfloor#3\right\rfloor} 
    {#2\lfloor#3#2\rfloor} 
}

\newcommand{\textb}[1]{\textcolor{black}{#1}}

\newcolumntype{C}[1]{>{\centering\let\newline\\\arraybackslash\hspace{0pt}}m{#1}}

\algrenewcommand\algorithmicrequire{\textbf{Input:}}
\algrenewcommand\algorithmicensure{\textbf{Output:}}

\acrodef{5G}{fifth generation}
\acrodef{3G}{third generation}
\acrodef{4G}{fourth generation}
\acrodef{3GPP}{3rd Generation Partnership Project}
\acrodef{AWGN}{additive white Gaussian noise}
\acrodef{F-BS}{fog-computing-based BS}
\acrodef{BS}{base station}
\acrodef{BBU}{baseband unit}
\acrodef{CA}{carrier aggregation}
\acrodef{CB}{carrier bonding}
\acrodef{CC}{carrier component}
\acrodef{CR}{cognitive radio}
\acrodef{CSI}{channel state information}
\acrodef{CCI}{co-channel interference}
\acrodef{CoMP}{coordinated multi-point}
\acrodef{HetNet}{heterogeneous network}
\acrodef{D2D}{device-to-device}
\acrodef{DSA}{dynamic spectrum access}
\acrodef{DL}{downlink}
\acrodef{EE}{energy efficiency}
\acrodef{FD}{full-duplex}
\acrodef{HD}{half-duplex}
\acrodef{IoT}{Internet-of-Things}
\acrodef{LED}{light emitting diode}
\acrodef{LOS}{line-of-sight}
\acrodef{mMTC}{massive machine-type communication}
\acrodef{MIMO}{multiple-input multiple-output}
\acrodef{mMIMO}{massive \acl{MIMO}}
\acrodef{mmWave}{millimeter wave}
\acrodef{NOMA}{non-orthogonal multiple access}
\acrodef{NLOS}{non line-of-sight}
\acrodef{NR}{New Radio}
\acrodef{OMA}{orthogonal multiple access}
\acrodef{OWC}{optical wireless communication}
\acrodef{PPP}{Poisson point process}
\acrodef{QoS}{quality-of-service}
\acrodef{RAN}{radio access network}
\acrodef{SA}{shared access}
\acrodef{SE}{spectral efficiency}
\acrodef{SC}{small-cell}
\acrodef{SBS}{small-cell \ac{BS}}
\acrodef{TVWS}{TV white space}
\acrodef{RRH}{remote radio head}
\acrodef{RF}{radio-frequency}
\acrodef{SINR}{signal-to-interference plus noise ratio}
\acrodef{SI}{self-interference}
\acrodef{SIC}{self-interference cancellation}
\acrodef{SuIC}{successive interference cancellation}
\acrodef{UL}{uplink}
\acrodef{UDenseNet}{ultra-dense network}
\acrodef{URLLC}{ultra-reliable low-latency communication}
\acrodef{VLC}{visible light communication}

\begin{document}

\title{All Technologies Work Together for Good: \\A Glance to Future Mobile Networks }
\author{\noindent \begingroup\centering{Animesh Yadav,~\IEEEmembership{Member,~IEEE}, and Octavia A. Dobre \IEEEmembership{Senior Member, IEEE}}\endgroup%
\thanks{
This work has been supported in part by the Natural Sciences and Engineering Research Council of Canada (NSERC),
through its Discovery program.

Animesh Yadav and Octavia A. Dobre are with the Faculty of Engineering and Applied Science, Memorial University, St. John's, NL, A1B 3X5, Canada (e-mail: \{animeshy, odobre\}@mun.ca).}
}


\maketitle
\begin{abstract}
The astounding capacity requirements of 5G have motivated researchers to investigate the feasibility of many potential technologies, such as massive multiple-input multiple-output, millimeter wave, full-duplex, non-orthogonal multiple access, carrier aggregation, cognitive radio, and network ultra-densification. The benefits and challenges of these technologies have been thoroughly studied either individually or in a combination of two or three.  It is not clear, however, whether all potential technologies operating together lead to fulfilling the requirements posed by 5G. This paper explores the potential benefits and challenges when all technologies coexist in an ultra-dense cellular environment. The sum rate of the network is investigated with respect to the increase in the number of small-cells and results show the capacity gains achieved by the coexistence.
\end{abstract}

\IEEEpeerreviewmaketitle{\noindent }

\section{Introduction}
\IEEEPARstart{I}{n} the last decade, the advancements in the very-large-scale integrated-circuit electronics have led to an advent of affordable smart handheld wireless devices, which are now within reach of millions of people. These devices are commonly used for Internet-based applications, such as online banking, e-commerce, social media, multimedia streaming, and online gaming besides the voice application. For Internet-based services, each device needs to connect to a wireless network. As a result, a high number of subscribers straightforwardly translates into high Internet data traffic in the wireless network. Additionally, the billions of \ac{IoT} devices, with applications to smart healthcare, smart homes and cities, smart electricity grids, etc., are expected to use the wireless network. According to Ericsson's  technical mobility report published in 2017, around 29 billion connected devices are forecast by 2022, of which around 18 \textb{billions} will be related to IoT. Hence, it is quite evident that one of the key challenges in future wireless networks is the unprecedented high volume of data traffic. 

Towards accomplishing the anticipated humongous volume of data traffic among other requirements, telecom industry leaders, standardization agencies, and academia are working on solutions for upcoming \ac{5G} of wireless network. Particularly, the third generation partnership project (3GPP) has initiated the \ac{5G} \ac{NR} standardization process. The 5G NR is not just about achieving 1000 times higher network \ac{SE}, but also about providing ubiquitous connectivity, guaranteed \ac{QoS}, low-cost and \ac{mMTC}, 100 times better \ac{EE}, support for 500 km/h mobility, \acl{URLLC}, and increased battery operating lifetime of devices \cite{Parkvall-Dahlman-Furuskar-Frenne-csm-2017}.

This article discusses solutions that have the capability of delivering the extreme capacity demand of the \ac{5G} and beyond networks. Broadly, techniques for capacity enhancement of a network can be classified into four classes: i) use of additional wider bandwidth; ii) spectrum reuse;  iii) use of new technologies; iv) efficient spectrum management. Within these classes, several potential technologies have been identified to achieve the goals setup for \ac{5G} networks. Among them, the \ac{UDenseNet} of \acp{SC}, \ac{FD} transceivers, \ac{mmWave} communications, \ac{mMIMO}, \ac{NOMA}, and \ac{DSA} technologies are the primary ones in achieving the goal of large capacity. 

Individually, each technology has great potential benefits. Some of these technologies complement each other when they are combined to work together, and can achieve higher performance gains compared to their individual performances. For example, the joint use of \ac{mmWave} and \ac{mMIMO} technologies provides tangible capacity enhancement \cite{Bogale-Le-VTM-2016}. The highly \textb{directional} beams are desirable for \ac{mmWave} communications, which can be obtained by using \ac{mMIMO}. Additionally, small-wavelengths at \ac{mmWave} frequencies help in realizing small-size antennas, which are essential for \ac{mMIMO} deployment.  Nevertheless, combining all these technologies is not trivial, and poses additional challenges. In this article, we study the benefits and challenges of simultaneously operating these technologies, and bring the following contributions:
\begin{itemize}
\item A comprehensive overview of the key enabling technologies for 5G and beyond wireless networks is provided and their fundamental features, advantages and disadvantages are introduced;
\item The state-of-the-art coexistence of two or three technologies is reviewed and a novel coexistence of all key technologies is envisaged for throughput enhancement in future wireless networks. Open challenges related to the coexistence are also discussed;
\item The impact of the number of \acp{SC} on the sum rate in coexistence scenarios is quantitatively studied using numerical simulations, and insights are presented.   
\end{itemize}

The rest of this article is organized as follows. In Section II, we summarize the fundamental features, benefits and challenges of key technologies. In Section III, we discuss the feasibility of the coexistence of these technologies for \ac{5G} and beyond networks. Open research issues are introduced in Section IV. In Section V, we present numerical simulation results for an \ac{UDenseNet}, where such technologies work together. Finally, the paper is concluded in Section VI.

\section{Key Enabling Technologies For 5G Networks And Beyond}
This section provides an overview of technologies that play a pivotal role for emerging 5G and beyond wireless networks. A brief summary of their fundamental features, advantages, and disadvantages is presented in Table~I.
\definecolor{LightCyan}{rgb}{0.88,1,1}
\newcolumntype{g}{>{\columncolor{LightCyan}}c}
\newcolumntype{b}{>{\columncolor{blue}}c}
\begin{table*}[ht]\label{long}
\caption{A brief summary of various key enabling technologies for 5G and beyond.}
\centering
 \rowcolors{3}{LightCyan!50}{white}
  \begin{tabular}{c|C{4cm}|C{5cm}|C{4cm}}
    \rowcolor{LightCyan!100}
    \hline
    \textbf{Technologies} & \textbf{Fundamental features} & \textbf{Advantages} & \textbf{Disadvantages}\\
    \hline
    \textbf{UDenseNets} & Amorphous deployment of hundreds of \acp{SC} in a macro-cell region, i.e., small ratio between the number of \acp{SC} and users & Large coverage, low power, high throughput, enables spatial frequency reuse & Inter-cell \acs{CCI}\\
    \textbf{mmWave} and \textbf{VLC} & Large amount of bandwidth & Both provide high \acs{SE};
    
     mmWave: enables the use of a large number of miniature-size antennas; 
    
    VLC: low-cost infrastructure, i.e., LEDs and inherited link security  & mmWave: high signal attenuation, short transmission distance; 
    
    VLC: low-modulation bandwidth and inter-symbol interference\\
    \textbf{mMIMO} & Use of hundreds of antennas for communication & Surplus degrees of freedom including higher \acs{SE} and \acs{EE}, low-power transmission, simple linear signal processing  & Requires a large number of \acs{RF} chains, as well as estimation of a large number of \acs{CSI} coefficients\\
    \textbf{FD} & Transmit and receive on the same time and frequency resource & Theoretically doubles the \acs{SE} without additional spectrum requirement & Self-interference and \acs{CCI}\\
    \textbf{NOMA} & Multiplex several users on the same time and frequency resource & Enhances \acs{SE} without additional spectrum requirement, better user fairness & Multiuser interference\\
    \textbf{DSA} & Utilization of underused spectrum & Improves spectrum utilization efficiency & Interference from secondary to primary users or networks\\
    \hline
  \end{tabular}
\end{table*} \phantom{\ac{BS}}

\subsection{Network Ultra-Densification}
The concept of cell densification has evolved from 4-5 macro-cell base stations (MBSs)/km$^2$ in the \acl{3G} wireless networks to about 8-10 micro-cell BSs/km$^2$ in the \acl{4G}. The primary aim of cell densification is to address the problem of capacity and coverage by spatial frequency-reusing and offloading the data traffic to the \acp{SC} \cite{Bhushan-Li-Malladi-Gilmore-Brenner-Damnjanovic-Sukhavasi-Patel-Geirhofer-mcom-15}. The macro-cells in the \acl{4G} have a smaller area than those in the \acl{3G}. Further, \acp{SC} bring users closer to the \ac{BS}, which reduces the access distance, and consequently, the path loss between them.  In \ac{5G}, the \ac{SC} area would be reduced further to support low-power transmissions, and hence, the cell density is expected to increase to 40-50 \acp{SBS}/km$^2$. However, the \acp{UDenseNet} increase the interference power levels, and consequently, the overall network performance might degrade due to inter-cell \ac{CCI}. Hence, interference mitigation is paramount for a network to achieve high \ac{SE}.  
 
\subsection{Millimeter-Wave and Optical Wireless Technologies }
It is evident that current wireless cellular networks are unable to cater the future throughput demand, as the existing frequency bands are limited and congested. A straightforward approach is to explore higher frequencies, which offer increased bandwidth. To this end, \ac{mmWave} bands ranging from 30-300 GHz, have been investigated for cellular purposes \cite{Rappaport-Others-ACCESS-2013}. At such frequencies, the main technical challenge is the severe attenuation due to path loss, shadowing, and blockage, along with the high energy consumption. The directional beams are used to combat the path loss and other losses.

Another promising approach is the \ac{OWC}, which enables the use of large bandwidths in optical bands such as infrared, visible light, and ultra-violet spectra. In particular, the visible light spectrum  band ranging from 400-800 THz can be used freely for cellular purposes. The \ac{VLC} exhibits inherited network security, and it is easier to integrate using existing lightening infrastructure, e.g., \acp{LED}. Since the visible light does not penetrate through opaque materials, \ac{VLC} is a suitable candidate for indoor environments. The main challenges in \ac{VLC} are the low-modulation bandwidth of \acp{LED}, inter-symbol interference and \ac{CCI}.

\subsection{Massive MIMO Technology}
The \acs{MIMO} technology has been well recognized for enhancing the \ac{SE}, \ac{EE}, and reliability of a wireless link. Due to the use of multiple antennas at both transmitter and receiver sides, a \acs{MIMO} system provides multiplexing and diversity gains. In a multi-user case, \acs{MIMO} enables communication of multiple users simultaneously on the same radio resource. On the other hand, by significantly increasing the number of antennas to tens or hundreds, i.e., \ac{mMIMO} system, the \ac{SE} and \ac{EE} can be significantly improved beyond those of the traditional \acs{MIMO} systems. Other advantages associated with \ac{mMIMO} are: simplified signal processing, such as zero-forcing transmitter and maximum ratio combining receiver; channel hardening effect; and low-power transmission due to high beamforming and array gains \cite{Rusek-Others-SPM-2013}.  The major difficulty in deploying \ac{mMIMO}  is the requirement of a large number of \ac{RF} chains and a large number of \ac{CSI} coefficient acquisition, as well as pilot contamination. A combination of digital and analog beamformers is an efficient solution to reduce the number of \ac{RF} chains. Various efficient hybrid beamforming structures are discussed in \cite{Molich-Others-CM-2017}.   
    
\subsection{Full-Duplex Technology}
Recently, the \ac{FD} technology has become an active research area because of its ability to double the SE of a network without requiring new spectrum. Hence, the \ac{FD} technology has the potential to contribute towards meeting the high capacity and low-latency \ac{5G} requirements. However, an \ac{FD} transceiver needs to cope with \ac{SI}, which suppresses the low-power received signal by the high-power transmitted signal. Several works have focused on developing efficient \ac{SI} cancellation techniques. The experimental results show that by applying a combination of passive and active cancellation methods, \ac{SI} can be made as low as the thermal noise level. The active methods include analog cancellation at the \ac{RF} stage and digital cancellation at the baseband stage. Another performance limiting factor is the \ac{CCI} between the \ac{DL} and \ac{UL} users, which can be reduced by employing optimal power allocation and scheduling schemes. 

\subsection{Non-Orthogonal Multiple Access}
Likewise \ac{FD} technology, \ac{NOMA} is another promising cost-effective technique that improves the \ac{SE} of a network without the need of additional bandwidth resources. Furthermore, it also ensures user fairness, which is an important metric from the perspective of network operators. In \ac{NOMA}, multiple users are multiplexed mainly in code- or power-domain and transmit over the same frequency and time resource. For example, sparse code multiple access is a code-domain \ac{NOMA} scheme, where the codewords of different users are superimposed non-orthogonally using sparse spreading, and multiuser detection and data recovery are jointly performed at the receiver side. On the other hand, in power-domain \ac{NOMA} \ac{DL} transmission with two users, a \ac{BS} superimposes the user data and allocates more transmit power to the user with worse channel conditions. Each user then performs the 
\ac{SuIC} to decode the information by eliminating the undesired interference \cite{Riaz-Avazov_Dobre-Kwak-COMST-2017}. On one hand, on a \ac{DL} channel, \ac{NOMA} outperforms the traditional \ac{OMA} scheme, in terms of both sum rate and fairness. On the other hand, when compared to the non-linear dirty-paper coding, \ac{NOMA} has lower computational complexity; however, at the cost of some performance loss.

It is worth mentioning that \ac{UL} \ac{NOMA} can support massive connectivity, which is highly desirable for \ac{mMTC} scenarios. Furthermore, \textb{in a grant-free UL access, latency reduction could be expected due to control signaling minimization.} 

\subsection{Dynamic Spectrum Access}
The previously discussed technologies yield increased \ac{SE}; however, they do not solve the problem of spectrum under-utilization. The \ac{DSA} technology provides a promising way that enables the use of underutilized frequency bands \cite{Tsiropoulos-Yadav_Zeng-Dobre-WCM-2017}. The \ac{DSA} techniques broadly fall into two categories: \text{i)} cognitive-inspired radio access, which allows the secondary networks to use temporarily or spatially unoccupied bands without harmful interference to the primary networks; \text{ii)} cooperative-inspired radio access, where the primary networks allow spectrum sharing, trading, and leasing to the secondary networks with some economic favor. For example, the DSA technology plays a key role in the seamless integration of the ad-hoc \ac{D2D} communications into the bandwidth scarce-network. The \ac{D2D} links are supposed to provide low-latency, low-power, and high peak data rate service between two network users.

\definecolor{LightCyan}{rgb}{0.88,1,1}
\newcolumntype{g}{>{\columncolor{LightCyan}}c}
\newcolumntype{b}{>{\columncolor{blue}}c}
\begin{table*}[h]\label{long-2}
\caption{A brief summary of the state-of-the-art combination of various key enabling technologies for 5G and beyond.}
\centering
 \rowcolors{3}{LightCyan!50}{white}
  \begin{tabular}{C{3.2cm}|C{6cm}|C{4.5cm}|C{2.5cm}}
    \rowcolor{LightCyan!100}
    \hline
    \textbf{Combinations} &\textbf{Coupling factors} & \textbf{Problems} & \textbf{Possible solutions}\\
    \hline
    \textbf{UDenseNets}  and \textbf{mmWave} \cite{Feng-Wang-Lin-Ge-Lu-Li-jsac-2017} & Large network throughput and \ac{mmWave} signals travel short distance due to large propagation loss; hence, they are a good fit for \acp{UDenseNet} & In an urban environments, \acs{mmWave} suffers further propagation loss due to shadowing and blockage; hence, \textb{it requires} highly \textb{directional} beams for transmission & \acs{mMIMO}\\
    \textbf{mmWave} and \textbf{mMIMO} \cite{Bogale-Le-VTM-2016} & \acs{mMIMO} helps in achieving the much needed highly directed beam for \acs{mmWave} operation, and the small wavelengths of \acs{mmWave} help accommodating a large number of miniature-size antennas in a small space & The digital beamforming architecture for \acs{mMIMO} requires a large number of \ac{RF} chains corresponding to each antenna, and unaffordable energy consumption & Hybrid beamforming architecture and beamspace \acs{MIMO}\\
    \textbf{mMIMO} and \textbf{NOMA} \cite{Wang-Dai-Wang-Ge-Zhou-jsac-2017} & Especially, beamspace \acs{MIMO} supports one user per \ac{RF} chain, which limits the number of users that the \acs{SBS} can support. Integrating beamspace \acs{MIMO} with \acs{NOMA} technique eliminates this limitation & Cannot support large number of users due to the decoding complexity of \acs{SuIC}, which scales with the number of users, as well as \textb{due} to the problem of error propagation in \acs{SuIC} \cite{Shin-Vaezi-Lee-Love-Poor-CM-2017}  & \textb{Group a small number of users per \acs{RF} chain} and apply \acs{FD} communication for each chain\\
    \textbf{mMIMO} and \textbf{FD} \cite{Shojaeifard-others-tcom-2017} \textbf{NOMA} and \textbf{FD} \cite{Sun-Ng-Ding-Schober-tcom-2017} & The highly directional beams generated by \acs{mMIMO} help reducing the LOS \acs{SI} at the \acs{FD} transceiver;
    
     NOMA and FD together improve \acs{SE}. NOMA improves user fairness & Huge surge of intra- and inter-cell \acs{CCI} due to FD communication& Efficient resource management and interference mitigation techniques \\ 
    \hline
  \end{tabular}
\end{table*}

\section{A Future Cellular Network Design}
This section discusses the future cellular network design, where all key enabling technologies coexist. A summary of a few state-of-the-art technologies that are combined with a perspective of improving the overall network throughput is given in Table~II.
\subsection{Coexistence of Key Enabling Technologies}
In addition to the cellular traffic, the next generation wireless network supports traffic from \ac{D2D} and \ac{IoT} communications as well. Hence, seamless coexistence of all key technologies is pivotal to achieve a large throughput in the \ac{5G} and beyond wireless networks. Moreover, from the perspective of the infrastructure of a macro-cell belonging to a \textb{fog-computing-based radio access network (F-RAN)} architecture, a significant difference that the upcoming future networks would bring is the massive deployment of several low-cost and low-power \acp{SBS}. Several fixed and mobile relays, as well as \textb{\acp{F-BS}} and a large number of \acp{RRH} will be used for an enhanced performance. The \textb{\acp{F-BS}} are fixed low-power \acp{BS}, which are connected \textb{to the core network either directly or through the pool of \acp{BBU} or through the MBS. The \textb{\ac{F-BS}'s} connections to the \ac{BBU} and MBS are performed using fronthaul links and optical fibers, respectively.} The MBS and \acp{RRH} are connected to the  \acp{BBU} for baseband processing \textb{via X2 interface and fronthaul, respectively}. The \textb{\acp{F-BS}} have extra signal processing functionalities to perform some tasks, e.g., cooperative \textb{radio signal processing and} interference management, locally rather than centrally at the \ac{BBU}. \textb{Local processing helps in reducing latency and burden on the fronthaul and \ac{BBU}.} A graphical representation of a macro-cell infrastructure and links is depicted in Fig.~1. For a seamless coverage, users closer to \acp{SBS} and \acp{RRH} are served by \acp{SBS} and \acp{RRH}, respectively.

To better understand the potential benefits and challenges posed by the co-existence of all key enabling technologies, we divide the discussion into wireless access and backhaul\textb{ing in the \ac{SC} networks}.
\subsection{Wireless Access \textb{in the \acp{SC}} Network}
\indent 1) \textit{UDenseNets, \ac{mmWave} and \ac{OWC}:} The amorphous and ultra-dense deployment of \acp{SC} reduces the access distance between the users and \ac{SBS}, and provides an unlimited user experience. The \acp{SC} can be deployed in both indoor and outdoor environments. However, using  the existing cellular bands for communication in these cells should be avoided as they travel far beyond the cell boundaries and cause severe inter-cell \ac{CCI}. The \ac{mmWave} bands are a good alternative as the signals travel shorter distance because of the large propagation loss, and hence, generate low inter-cell \ac{CCI}. Moreover, for an indoor case, the spectrum from the optical bands is an option as well. For instance, in a room or a corridor of a building or a house, there can be a \ac{VLC}-\ac{SC} that serves several users without creating inter-cell \ac{CCI}. Hence, the \ac{mmWave} and \ac{OWC} technologies complement well with the network densification \cite{Feng-Wang-Lin-Ge-Lu-Li-jsac-2017}. In an outdoor urban environment, \ac{mmWave} is a viable option; however, the users in a \ac{mmWave}-\acp{SC} achieve smaller \acl{SINR} due to the blockage effect.
 
\indent 2) \textit{UDenseNets, \ac{mmWave}, \ac{OWC} and \ac{mMIMO}:} In order to improve the data rate performance of the \ac{mmWave}-\ac{SC} users and cope with the large propagation loss, the \ac{mmWave} signal needs to be highly directional, which can be achieved either by using a large number of antennas or directional antennas. The small wavelength at \ac{mmWave} frequencies allows packing a massive amount of miniature-size antennas to form an antenna array for transceivers. Additionally, the massive amount of antennas not only boosts up the received signal power via array gain, but also enhances the \ac{SE} of the network using multiuser \acs{MIMO} techniques.

However, the \ac{mmWave} and \ac{mMIMO} integration requires a large number of \ac{RF} chains that accounts for approximately $70\%$ of the total transceiver energy consumption \cite{Rusek-Others-SPM-2013}. Hence, this particular integration is economically prohibitive, and energy and cost-efficient solutions have been explored. A fully analog and hybrid (analog-digital) beamforming architectures have been \textb{investigated}. The fully analog architecture requires only one \ac{RF} chain with an analog circuit for adjusting the phases of signals. As such, this architecture supports a single transmit beam, which makes difficult to adjust the beam to the channel conditions, and results in considerable performance loss. On the other hand, the hybrid beamforming architecture supports a minimum number of \ac{RF} chains, which helps in reaping the benefits of multiple antennas, and offers a reasonable trade-off between cost and performance \cite{Molich-Others-CM-2017}. A different alternative solution which has recently surfaced is the beamspace \acs{MIMO} technique; this significantly reduces both the number of required \ac{RF} chains and the energy consumption \cite{Brady-Others-AnP-2013}. However, in the beamspace \acs{MIMO}, each \ac{RF} chain supports only one user at the same time-frequency resource, which is its fundamental limitation. Thus, the number of users supported by this scheme cannot exceed the number of \ac{RF} chains.
 
\indent 3) \textit{UDenseNets, \ac{mmWave}, \ac{OWC}, \ac{mMIMO}, and \ac{NOMA}:} When further integrating \ac{NOMA} with \ac{mmWave} and \ac{mMIMO} technologies, the limit of one user per \ac{RF} chain can be overcome through the multiplexing capability of \ac{NOMA}. However, in \ac{DL} power-domain \ac{NOMA}, at the user end, the decoding complexity of \ac{SuIC} scales with the number of users \cite{Zeng-Yadav-Tsir0poulos-Dobre-Poor-jsac-2017,Shin-Vaezi-Lee-Love-Poor-CM-2017}. Therefore, a smaller number of users, generally two users, are grouped per \ac{RF} chain. 
 
\indent 4) \textit{UDenseNets, \ac{mmWave}, \ac{OWC}, \ac{mMIMO}, \ac{NOMA}, and \ac{FD}:} Furthermore, by allowing each \ac{SBS} to operate in \ac{FD} mode, an \ac{RF} chain can serve users both in the \ac{DL} and \ac{UL} channels, which results in higher \ac{SE}. Theoretically, \ac{FD} communication doubles the \ac{SE} of the network when there is zero \ac{SI} and \ac{CCI}. As previously mentioned, the \ac{mmWave} technology enables the formation of antenna arrays by using massive miniature-size antenna elements that are arranged very close to each other. This arrangement leads to poor passive isolation between the transmit and receive antennas and causes higher \ac{SI}. Nevertheless, the beamforming achieved with the antenna array helps to overcome the problem of passive isolation. Moreover, highly directional signals have only low-power reflected \ac{NLOS} components that contribute towards the \ac{SI}.
 \begin{figure*}[t]
\centering
\includegraphics[width=2\columnwidth]{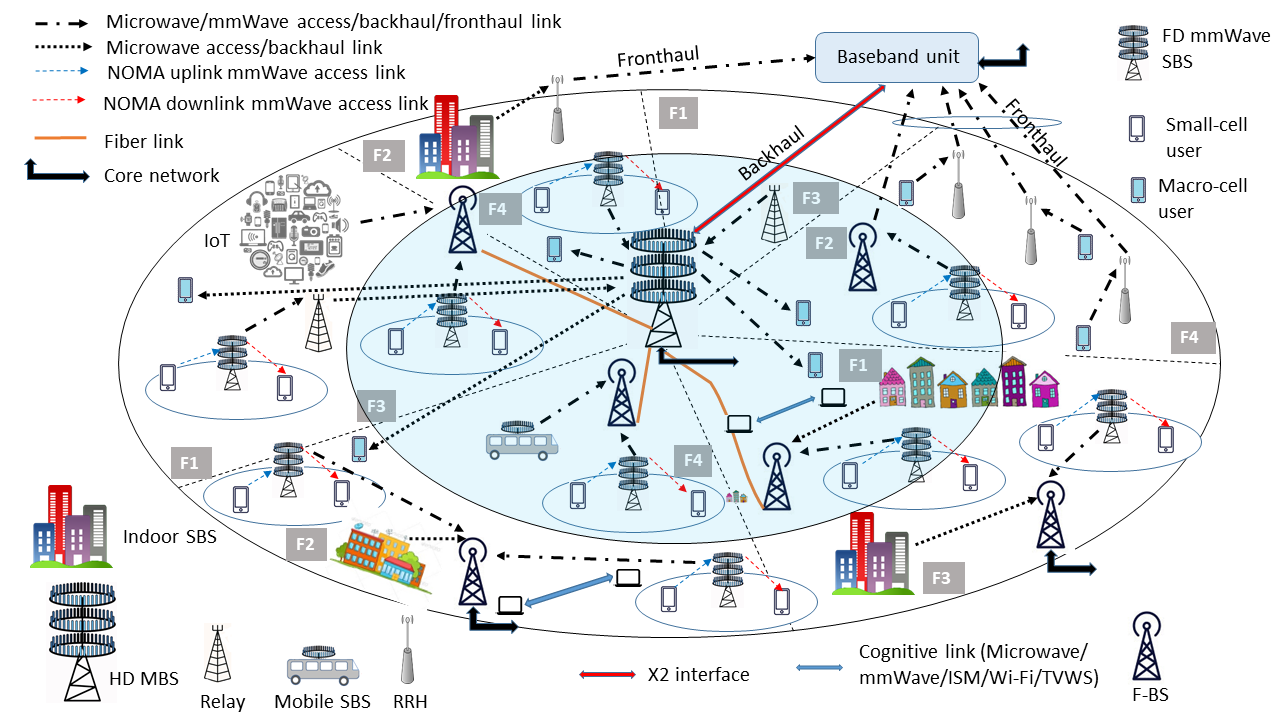}
\caption{The wireless access, backhaul and fronthaul links in a macro-cell belonging to an F-RAN architecture; the macro-cell has dense deployment of SCs, and uses technologies like mmWave, mMIMO, NOMA, and FD.}
\label{Spectral_Efficiency_Comparision}
\end{figure*} \label{access_network}
\subsection{Backhaul\textb{ing in the \acp{SC}} Network}
The high throughput supported by the access links is infeasible without high capacity backhaul links \textb{ for the \acp{SC}}. With the massive increase in the \ac{SC} deployment, a high capacity and reliable wired optical fiber backhaul connectivity for every \ac{SC} is uneconomical and inefficient. Evidently, the wireless backhaul is a suitable solution. However, the development of efficient wireless backhaul solutions is challenging because of the link capacity bottleneck, large network dimensionality, high energy consumption, and interference. 

To overcome the challenge of the capacity bottleneck of backhaul links, the spectrum in \ac{mmWave} bands can provide an appropriate solution. For instance, the \acp{SC}, which are in close proximity of the MBS can use the spectrum from the V- and D-bands without much restrictions. The close proximity region is the shaded area in Fig.~1, and is referred to as the close-zone. On the other hand, the \acp{SC} which are far from the MBS can transmit their backhaul data by using: i) dedicated relays; ii) dedicated \textb{\acp{F-BS}}; iii) the sub-6 GHz, beyond 6 GHz and licensed \ac{SA} spectrum from \ac{TVWS}, radar, and satellite bands. The far region is referred to as the far-zone, which is essentially the region between the boundaries of the close-zone and macro-cell in Fig.~1.

The dedicated relays for backhauling represent a potential solution over deploying the optical fiber cables; however, because of the large network size, this is not efficient in terms of overall capital expenditure and end-to-end delays. Hence, complete dependence on relays is not feasible. Consequently, fixed \textb{\acp{F-BS}} are expected to be deployed within a coverage area; they basically relay the aggregated backhaul traffic from several \acp{SBS} \textb{to core network either directly or through \ac{BBU} or MBS}. Since the \textb{\acp{F-BS}} are fixed, and less in number compared with the number of \acp{SC}, their connectivity \textb{to the MBS} via optical fiber is viable. Further, the \textb{\acp{F-BS}} can communicate among themselves to exchange information about the serving \acp{SC} for performing \ac{CoMP} operations to managing the inter-cell \ac{CCI}, and thus, improve the network capacity and cell edge user throughput.   

The next-generation cellular networks can also employ \ac{CA} and \ac{CB} techniques for creating a wider bandwidth by using the spectrum from licensed, unlicenced and \ac{SA} bands. \ac{CA} techniques can aggregate multiple contiguous and/or non-contiguous \acp{CC} belonging to intra- and/or inter-bands. \ac{CB} techniques are used to combine contiguous intra-band \acp{CC} in a single wider band. The licensed and \ac{SA} spectrum available to use is from sub-6 GHz, including TVWS (600-800 MHz), radar (960-1400, 2700-3650, 5-5.850 GHz), satellite's L band, as well as from beyond-6 GHz including spectrum from \ac{mmWave}, satellite's (S/C bands), and optical bands. The \ac{CA}/\ac{CB} in these bands can be enabled through \ac{DSA} techniques such as sensing, active scanning, and geo-location databases and beaconing. The spectrum from \ac{TVWS}, radar, and satellite belongs to SA bands, and hence, its usage in the cellular network is performed either via a cognitive underlay or overlay approach. Based on the backhaul link-distance, the MBS can select the appropriate spectrum band for backhaul communication. A distance-based spectrum usage scheme that relies on the \ac{CA}/\ac{CB} techniques is shown in Fig.~1. The \acp{SBS} and relays located in the \textb{close-zone} tend to use frequencies from \ac{mmWave}, while they tend to use frequencies from TVWS, radar, and satellite bands when located in the \textb{far-zone}.  

\begin{figure}[t]
\centering
\includegraphics[width=1\columnwidth]{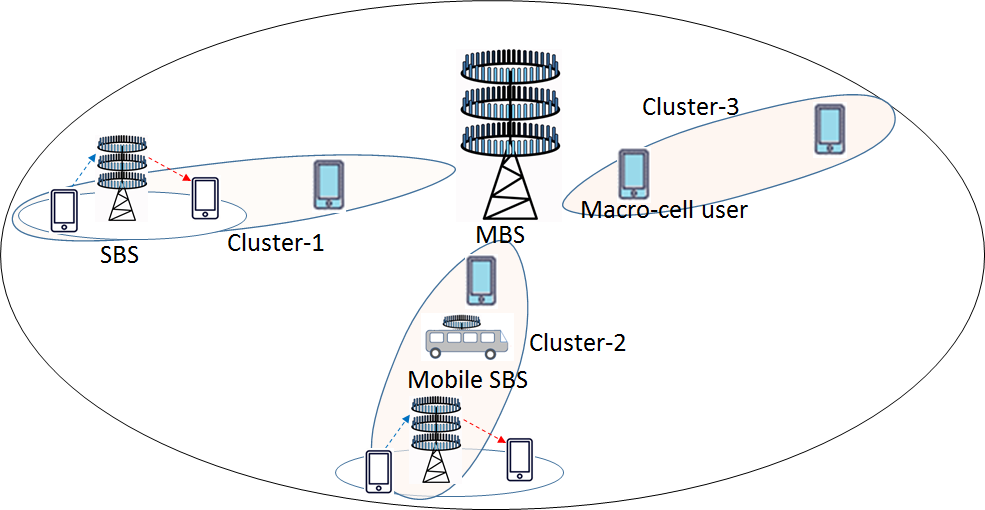}
\caption{The MBS serves several SBSs and macro-cell users by grouping them in three clusters.}
\label{Spectral_Efficiency_Comparision}
\end{figure} \label{NOMA_MBS}
Furthermore, the available spectrum can be utilized efficiently by allowing the \acp{SBS} to operate in \ac{FD} mode, i.e., simultaneous transmission to the users and backhaul transmission to the relay or \textb{\ac{F-BS}} or MBS. Additionally, by applying \ac{DL} power-domain \ac{NOMA} at the MBS to superimpose the data of two or more \acp{SBS} or of macro-cell users and \acp{SBS}, improves the backhaul capacity; such a scenario is illustrated in Fig.~2.

\section{Open Challenges}
It is certain from the above discussion that the coexistence of all key technologies would significantly improve the capacity of the next generation wireless networks. However, this also brings challenges that cannot be overlooked; some are mentioned as follows:

\indent \textit{Interference management:} A major challenge is the intra- and inter-cell \ac{CCI} in both \ac{mmWave} and microwave frequencies. The highly directional \acs{LOS} beams are essential at \ac{mmWave} frequencies in the \ac{SC} for improving the overall cell throughput; however, this decreases the cell-edge user's throughput due to inter-cell \ac{CCI}. On the other hand, at microwave frequencies, both \acs{LOS} and \ac{NLOS} reflected beams create a strong \ac{CCI} to the users both in and out of the \ac{SC}. The interference level is more profound in an amorphously deployed \acp{UDenseNet}. Further, the \ac{FD} and \ac{NOMA} links create additional inter-cell \ac{CCI} and limit the performance. Hence, interference management is critical and plays a vital role in the seamless coexistence of key technologies. The \ac{CoMP} techniques for interference and coordinated resource management are certainly helpful here; however, they require large amount of \ac{CSI}. The distributed techniques would provide reasonable solutions with some performance loss.

\indent \textit{Joint SBS and RRH selection optimization:} In \acp{UDenseNet}, a user would be in the close proximity of many cells. The \acp{SC} in the neighbourhood of a user could be either dominant interferers or strong servers. This depends on the association scheme and the coordination among the neighboring \acp{SC}. Furthermore, the performance of the overall network is heavily restricted by the backhaul links. As \ac{FD} and \ac{NOMA} techniques enable even more users to be served by an \ac{SC}, the burden on the backhaul links increases. Hence, it is imperative for each \ac{SBS} to select an appropriate \textb{\ac{F-BS}} or relay. Similarly, for the users associated to \acp{RRH}, to select the  \ac{RRH} that has a larger link capacity to the BBU. However, a joint user association to the \ac{SBS} \textb{and/or} \ac{RRH}, and selection of \textb{\ac{F-BS}} or relay scheme needs to be investigated; such a solution can further improve the overall network throughput performance.

\indent \textit{Green network architecture:} Owing to the network densification in the 5G and beyond networks, the \acp{SC} can dynamically adjust their power based on the current needs. To this end, machine learning algorithms are useful as they learn the traffic patterns of individual \ac{SC} to switch the \acp{SBS} on or off. The on/off scheme for the \acp{SBS} offers significant improvement in the energy utilization of the network. 
\section{Performance Studies}
In this section, we demonstrate the access links throughput gain achieved by implementing \ac{FD} and \ac{NOMA} at \acp{SC} in an \ac{UDenseNet}, which operates in mmWave band and employs mMIMO technology. We consider the macro-cell scenario as depicted in Fig.~1. In order to avoid the interference and improve the frequency reuse, the macro-cell area is partitioned into close- and far-zones. Each such zone is further sectored in six sub-regions. Four orthogonal mmWave bands, denoted as F1, F2, F3 and F4, are used in these sub-regions. Since the mmWave signal suffers huge propagation attenuations, the allocation of the set of four mmWave bands shown in Fig.~1 is good enough to avoid any inter-region interference. In simulations, we consider only one such sub-region scenario. In this sub-region, all the \acp{SBS} use the same frequency band for access links. 

The locations of \acp{SC} is modeled via a \acl{PPP} with density $\lambda_s$. The users are uniformly distributed in the macro-cell region. \textb{The mmWave channels between the users to the SBSs are modelled as describe in \cite{Wang-Dai-Wang-Ge-Zhou-jsac-2017}.} A summary of simulation parameters and their values is provided in Table~III. 

\definecolor{LightCyan}{rgb}{0.88,1,1}
\newcolumntype{g}{>{\columncolor{LightCyan}}c}
\newcolumntype{b}{>{\columncolor{blue}}c}
\begin{table}[ht]\label{sim_param}
\caption{Simulation parameters and their values.}
\centering
 \rowcolors{3}{LightCyan!50}{white}
  \begin{tabular}{m{3cm}|c}
    \rowcolor{LightCyan!100}
    \hline
    \textbf{Parameter} & \textbf{Values}\\
    \hline
    Transmit power (dBm) & SBS: 24, User: 20\\
    Noise and residual SI powers (dBm) & $\sigma_n^2$ = -104, and $\sigma^2_{\text{SI}}$ = -110\\
    Carrier frequency and system bandwidth  & $f_c$ = 28 GHz and 100 MHz \\
    NOMA power coefficients (weak and strong users) & 0.7 and 0.3 \\
    OMA power coefficients (weak and strong users) & 0.5 and 0.5 \\
    Number of transmitting antennas at SBS and user & 64 and 32\\
    Number of DL and UL users & 2 and 2\\
    Macro-cell and SC radius (in meters) &  500 and 100\\
    Path loss (in dB), where $d \geq$ 1 meter and $c$ is the speed of light & $\begin{array}{l} \text{LOS}: 20\log_{10}(4\pi f_c/c) + 20.1\log_{10}(d)\\
 \text{NLOS}: 20\log_{10}(4\pi f_c/c) + 34.0\log_{10}(d)\end{array}$ \\
    \hline
  \end{tabular}
\end{table}

\begin{figure}[t]
\centering
\includegraphics[width=1\columnwidth]{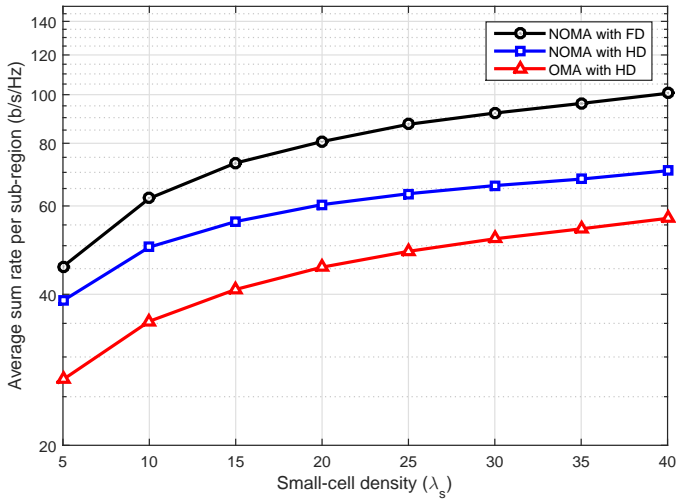}
\caption{A comparison of the sum rate of an \ac{UDenseNet} employing OMA-HD, NOMA-HD and NOMA-FD schemes, respectively.}
\label{Spectral_Efficiency_Comparision}
\end{figure} \label{NOMA_MBS}

We consider three cases for the \acp{SBS}, when they employ: i) \ac{OMA} and \ac{HD} schemes; ii) \ac{NOMA} and \ac{HD} schemes; and iii) \ac{NOMA} and \ac{FD} schemes. The HD and OMA are the traditional technologies, and hence, they are also plotted for gauging the performance over the NOMA and FD technologies. The beamspaced MIMO scheme \cite{Brady-Others-AnP-2013} is employed to reduce the number of \ac{RF} chains. In each \ac{SC}, on one hand, two \ac{DL} users are served by one \ac{RF} chain by multiplexing them using the \ac{NOMA} technique in a way similar to the one employed in \cite{Wang-Dai-Wang-Ge-Zhou-jsac-2017}. On the other hand, two \ac{UL} users are detected at the \ac{SBS} using the \ac{SuIC} technique. The intra- and inter-cell \acp{CCI} are considered as noise in the sum rate calculations. Further, we assume that for the \ac{FD} mode of operation, there are two \ac{DL} and \ac{UL} users in each \ac{SC} scheduled for communication. Contrastingly, for the \ac{HD} mode of operation, only two \ac{DL} users are assumed to be scheduled.

 Fig.~3 illustrates the average sum rate per sub-region of the network versus the density of \acp{SC} for all three cases. The results are obtained for 1000 random topologies. The sum rate is calculated using the Shannon's capacity formula. It can be observed that when both \ac{NOMA} and \ac{FD} are employed, a higher sum rate is achieved. In addition, the sum rate in all cases increases with the number of \acp{SC} and saturates later on due to high inter-cell \ac{CCI}. 
\section{Conclusion}

This article discusses important key enabling technologies, such as network densification, \ac{mmWave}, \ac{mMIMO}, \ac{NOMA}, \ac{FD}, and \ac{DSA}, which are identified as solution providers for  the capacity crunch problem in 5G and beyond wireless networks. To date, these technologies have been investigated either alone or in a combination of two or three together. This article envisages the integration of all technologies and discusses the overall benefits and challenges of their coexistence in wireless access and backhaul links. The sum rate of the network is investigated with respect to the increase in the number of \acp{SC}, and results show the capacity gains achieved by such coexistence. Lastly, this article paves the way for attaining better network capacity gains if the challenges are addressed efficiently.             
\renewcommand{\baselinestretch}{1}
\bibliographystyle{IEEEtran}

\section*{Biographies}
\vspace*{-3\baselineskip}
\begin{IEEEbiographynophoto}{\textsc{Animesh Yadav}}
 (animeshy@mun.ca) is a Research Associate at Memorial University, Canada. Previously, he worked as a Postdoctoral Research Fellow and Research Scientist at UQAM, Canada, and CWC at University of Oulu, Finland, respectively. During 2003-07, he worked as a Software Specialist at iGate Global Solutions Ltd., India and Elektrobit Oy, Finland. He received the master's and Ph.D. degree from Indian Institute of Technology (IIT) Roorkee, India and University of Oulu, Finland, respectively. He is the recipient of the best paper awards at IEEE WiMOB-2016 and IWCMC-2017. His research interests include enabling technologies for future wireless networks and green communications. \vfill
\end{IEEEbiographynophoto}
\vspace*{-2\baselineskip}

\begin{IEEEbiographynophoto}{\textsc{Octavia A. Dobre}}
 (odobre@mun.ca) is a Professor and Research Chair at Memorial University, Canada. In 2013 she was a Visiting Professor at Massachusetts Institute of Technology, USA, and University of Brest, France. Previously, she was with New Jersey Institute of Technology, USA, and Polytechnic Institute of Technology, Romania. She was the recipient of a Royal Society scholarship and a Fulbright fellowship. Her research interests include enabling technologies for 5G, cognitive radio systems, blind signal identification and parameter estimation techniques, as well as optical and underwater communications. She has authored around 200 referred journal and conference papers in these areas. Dr. Dobre is the EiC of the IEEE Communications Letters. She has served as editor for various prestigious journals, and technical co-chair of different international conferences, such as IEEE ICC and Globecom. She is a member-at-large of the Administrative Committee of the IEEE Instrumentation and Measurement Society and served as chair and co-chair of different technical committees. Dr. Dobre is a Fellow of the Engineering Institute of Canada.\vfill
\end{IEEEbiographynophoto}

\end{document}